\documentclass{cernrep}
\usepackage{graphicx,here}

\def\be{\begin{equation}}
\def\ee{\end{equation}}

\begin{document}

\title{STUDY OF THE COINCIDENCES BETWEEN THE
GRAVITATIONAL WAVE DETECTORS EXPLORER AND NAUTILUS IN THE YEAR 2001}

\author{
P. Astone$^1$,D. Babusci$^2$, M. Bassan$^3$, P. Bonifazi$^4$,
P. Carelli$^5$, G. Cavallari$^6$,\\
E. Coccia$^3$, C. Cosmelli$^7$, S.D'Antonio$^3$, V. Fafone$^2$,
G.Federici$^7$, S.Frasca$^7$,G. Giordano$^2$,\\ A. Marini$^2$,
Y. Minenkov$^3$, I. Modena$^3$, G. Modestino$^2$,
A. Moleti$^3$, G. V. Pallottino$^7$,\\ G. Pizzella$^8$,
L.Quintieri$^2$, A.Rocchi$^3$, F. Ronga$^2$, R. Terenzi$^4$, 
G.Torrioli$^9$, M. Visco$^4$\\
$~$\\
$~$
}

\vskip 0.1 in

\institute{
{\it ${}^{1)}$ Istituto Nazionale di Fisica Nucleare INFN, Rome}\\
{\it ${}^{2)}$ Istituto Nazionale di Fisica Nucleare INFN, Frascati}\\
{\it ${}^{3)}$ University of Rome "Tor Vergata" and INFN, Rome}\\
{\it ${}^{4)}$ IFSI-CNR and INFN, Rome}\\
{\it ${}^{5)}$ University of L'Aquila and INFN, Rome}\\
{\it ${}^{6)}$ CERN, Geneva, Switzerland}\\
{\it ${}^{7)}$ University of Rome "La Sapienza" and INFN, Rome}\\
{\it ${}^{8)}$ University of Rome "Tor Vergata" and INFN, Frascati}\\
{\it ${}^{9)}$ IESS-CNR, Rome}
}

\maketitle
 
\begin{abstract}
We report the result from a search for bursts of gravitational waves
using data collected by the cryogenic resonant detectors EXPLORER and NAUTILUS
during the year 2001, for a total measuring time of 90 days.
With these data we repeated the coincidence search performed
on the 1998 data (which showed a small coincidence excess) applying
data analysis algorithms based on known physical characteristics of
the detectors.
With the 2001 data a new interesting coincidence excess is found
when the detectors are favorably oriented with respect to the Galactic Disk.

\end{abstract}


\section{Introduction}
Cryogenic gravitational wave (GW) antennas entered into long term data
taking operation in 1990 (EXPLORER~\cite{long}), in 1991
 (ALLEGRO~\cite{alle}), in 1993 (NIOBE~\cite{NIOBE}),
in 1994 (NAUTILUS~\cite{naut}) and in 1997
(AURIGA~\cite{auri}), with gradual performance improvements over the years.

Analysis of the data taken in coincidence among all
cryogenic resonant detectors in operation during the years 1997 and 1998
was performed~\cite{5barre}.
No coincidence excess was found above background using
the event lists produced under the protocol of the
International Gravitational Event Collaboration (IGEC), among
the groups ALLEGRO, AURIGA, EXPLORER / NAUTILUS and NIOBE.

Later \cite{astone2001}, a coincidence search between the data of EXPLORER
and NAUTILUS was carried out by introducing in the data analysis 
considerations based on physical
characteristics of the detectors: the event energy and the directionality.
The result was a small coincidence excess when the detectors were
favorably oriented with respect to the Galactic Centre.

Here we extend our analysis to new data
obtained in the year 2001, when both EXPLORER and NAUTILUS were
operating at their best sensitivity,
using the same procedures applied for the previous analysis \cite{astone2001}.
As previously done in ref. \cite{astone2001} we shall sometimes use
the word $probability$,
although we are well aware that its significance might be jeopardized
by any possible data selection. With this $proviso$ we shall use
probability estimations in comparing different experimental conditions.

\section{Experimental data}

The resonant mass GW detectors NAUTILUS, operating at the INFN Frascati 
Laboratory, and EXPLORER, operating at CERN, both consist of an
Aluminium 2270 kg bar cooled to very low temperatures. 
A resonant transducer converts the mechanical
oscillations into an electrical signal and is followed by a
dcSQUID electronic amplifier. The bar and
the resonant transducer form a coupled oscillator
system with two resonant modes.

With respect to the year 1998 the following changes were made
in the set up of the detectors: the NAUTILUS detector operated 
at a thermodynamic temperature of 1.5 K instead of 0.14 K;
 the EXPLORER detector was 
equipped with a new transducer providing a larger bandwidth and consequently
enhanced sensitivity. The characteristics of the two detectors are given
in the Table \ref{dire}.

\begin{table}
\centering
\caption{
Main characteristics of the two detectors in the year 2001. The axes of
the two detectors are aligned to within a few degrees of one other,
the chance of coincidence detection thus being maximized. The pulse sensitivity
for both detectors is of the order of $h\sim 4~10^{-19}$ for 1 ms bursts.
}
\vskip 0.1 in
\begin{tabular}{|c|c|c|c|c|c|c|c|}
\hline
detector&latitude&longitude&azimuth&mass&frequencies&temperature&bandwidth\\
&&&&kg&Hz&K&Hz\\
\hline
EXPLORER&46.45 N&6.20 E&$39^o$ E&2270&904.7&2.6&$\sim 9$\\
&&&&&921.3&&\\
NAUTILUS&41.82 N&12.67 E&$44^o$ E&2270&906.97&1.5&$\sim 0.4$\\
&&&&&922.46&&\\
\hline
\end{tabular}
\label{dire}
\end{table}

The data, sampled at intervals of 12.8 ms for NAUTILUS and of
6.4 ms for EXPLORER,
are filtered with an adaptive filter matched to delta-like signals
for the detection of short bursts \cite{fast}. This search for bursts
is suitable for any transient GW which shows a nearly flat Fourier
spectrum at the two resonant frequencies of each detector. The metric
perturbation $h$ can either be a millisecond pulse, a signal made
by a few millisecond cycles, or a signal sweeping in frequency
through the detector resonances. This search is therefore sensitive to 
different kinds of GW sources, such as a stellar gravitational
collapse, the last stable orbits of an inspiraling neutron star
or black hole binary, its merging and its final ringdown.

Let $x(t)$ be the filtered output of the detector.
This quantity is normalized, using the detector calibration,
such that its square gives the energy innovation of the oscillation
for each sample, expressed in kelvin units.

For well behaved noise due only to the thermal
motion of the oscillators and to the electronic noise of the amplifier,
the distribution of $x(t)$ is normal with zero mean.
Its variance (average value of the square of $x(t)$)
is called $effective~temperature$ and is indicated with $T_{eff}$. The
distribution of $x(t)$ is
\be
f(x)=\frac{1}{\sqrt{2\pi T_{eff}}}e^{-\frac{x^2}{2T_{eff}}}
\label{normal}
\ee
In order to extract from the filtered data sequence $events$ to be
analyzed we set a threshold in terms of a critical ratio defined by
\be
 CR=\frac{|x|-\overline{|x|}}{\sigma(|x|)}
\label{creq}
\ee
where $\sigma(|x|)$ is the standard deviation of $|x|$  and
 $\overline{|x|}$ the moving average, computed over the preceeding ten minutes.

The threshold is set at CR=6 in order to obtain,
in the presence of thermal and electronic noise alone,
a reasonable number of events per day (see ref.\cite{astone2001}).
 This threshold corresponds to energy
$E_t=19.5~ T_{eff}$. When $|x|$ goes above the threshold, its
time behaviour is considered until it falls back below the threshold for
longer than three seconds. The maximum amplitude and its occurrence time
define the $event$.

The searched $events$ are the ones that are due to a combination of a GW signal
of energy $E_s$  and of the noise.
The theoretical probability to detect a signal with a given
signal to noise ratio $R_s=\frac{E_s}{T_{eff}}$,
in the presence of a well behaved Gaussian noise is \cite{papoulis}
\be
probability(R_s)=\int_{R_t}^{\infty} \frac {1}{\sqrt{2 \pi R_e}}
e^{-\frac{(R_s+R_e)}{2}}
cosh(\sqrt{R_e\cdot R_s})dR_e
\label{papou}
\ee
where $R_e$ is the signal to noise ratio for the event
and $R_t=\frac{E_t}{T_{eff}}=19.5$
for the EXPLORER and NAUTILUS detectors.

\begin{figure}
 \vspace{9.0cm}
\includegraphics{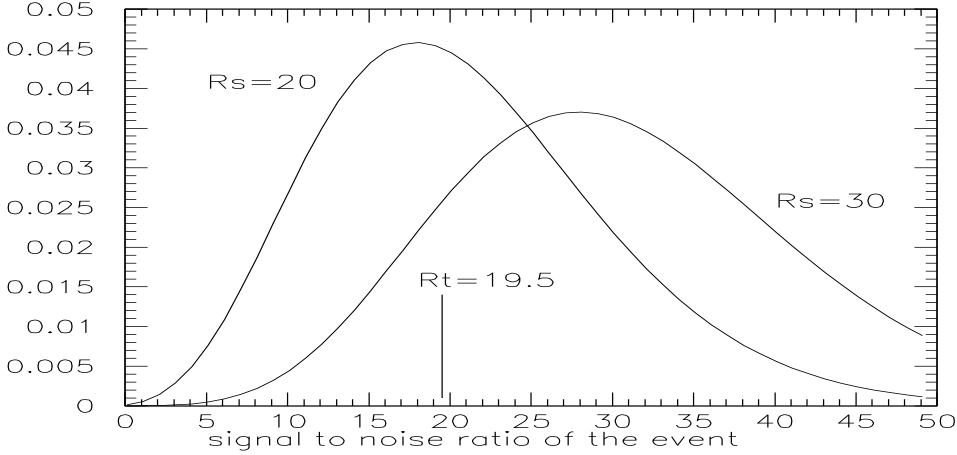}
 \caption{
Differential probability that the event has the signal-to-noise
ratio shown on the abscissa when the signal has $R_s=20$ 
(near the threshold $R_t=19.5$) and $R_s=30$.
        \label{fig1} }
\end{figure}

The behaviour of the integrand is shown in fig. \ref{fig1}.
This figure shows the spread of the event energy due to noise for a given
$R_s$ of the signal. It shows that signals with $R_s=20$ (near the
threshold)
have a probability of about 50\% not to be detected, and signals with
$R_s=30$, rather larger than the threshold, still have a probability of
near 15\% not to be detected. The distinction between the two concepts,
$signal$ and $event$, is essential for our analysis, as discussed
in ref \cite{astone2001}.

Computation of the GW amplitude $h$ from the energy signal $E_s$
requires a model for the signal shape. A conventionally chosen shape
is a short pulse lasting a time of $\tau_g$, resulting (for optimal
orientation, see later) in the relationship
\be
h=\frac{1}{4Lf^2} \frac{1}{\tau_g} \sqrt{ \frac{kE_s}{M}}
\label{he}
\ee
where $f$ is the resonance frequency, L and M the length and the
mass of the bar and $\tau_g$ is conventionally assumed equal to 1 ms
(for instance, for $E_s=1~mK$ we have $h=2.5~10^{-19}$).

\section{Data selection}

All events which are in coincidence
within a time window of $\pm 5~s$ with signals observed by a seismometer
are eliminated. This criterion cuts about $8\%$ of the events.

It is observed
that the experimental data are affected by non gaussian noise which,
 in some cases,
cannot be observed with any other auxiliary detector.
 Thus a strategy is needed to select periods during which the detectors
operate in a satisfactory way, as discussed in the following paragraphs.

\begin{figure}
 \vspace{9.0cm}
\includegraphics{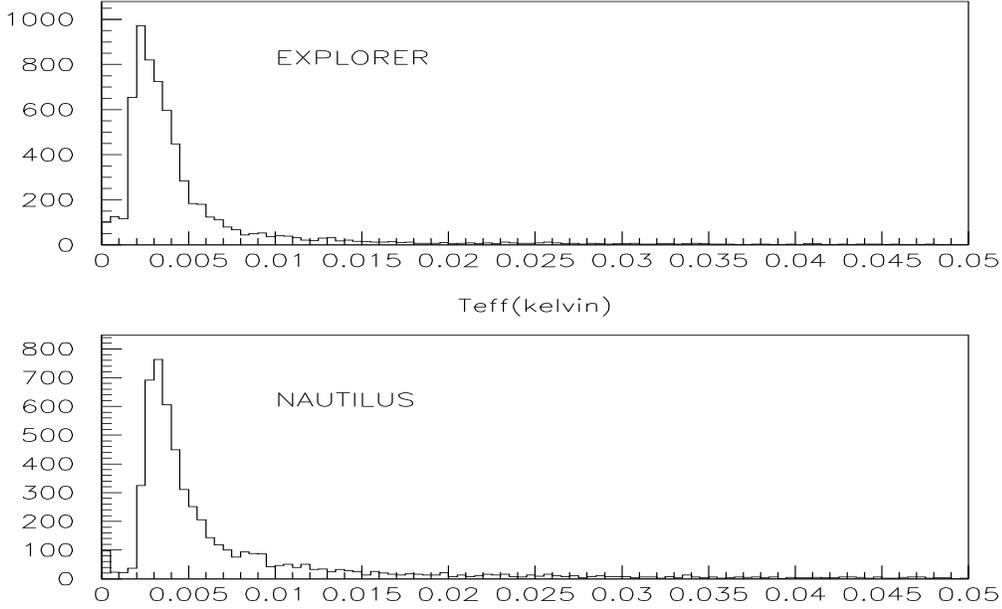}
 \caption{
The distributions of the hourly averages of
 $T_{eff}$ in kelvin units for EXPLORER and NAUTILUS.
We accept only the time periods with hourly averages 
$\overline{T_{eff}}~\leq~10~mK$.
        \label{fig2} }
\end{figure}

\begin{figure}
 \vspace{9.0cm}
\includegraphics{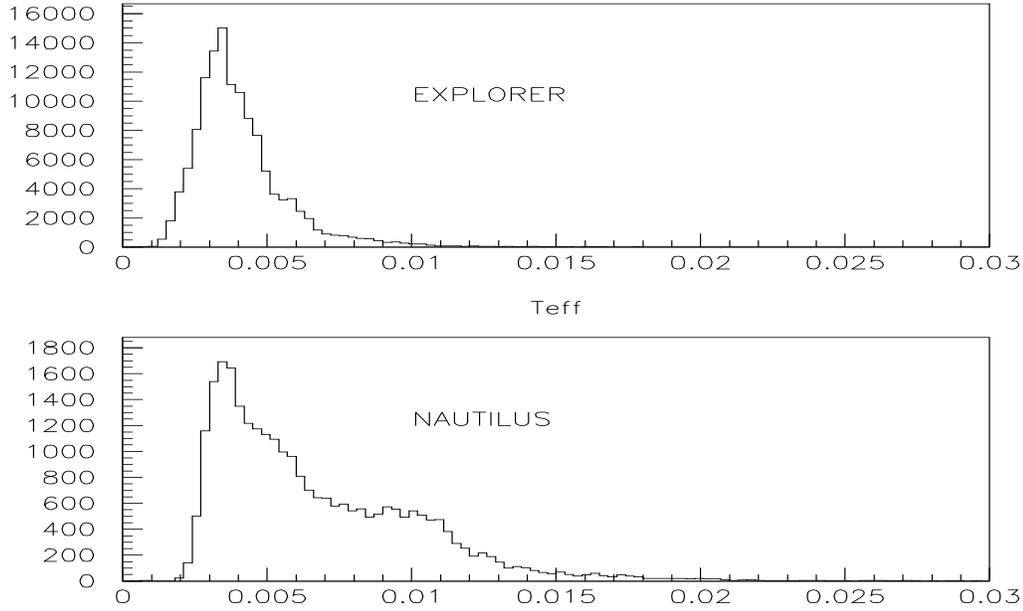}
 \caption{
The distributions of the ten minute averages of $T_{eff}$, before
each event, in kelvin
units for the EXPLORER and NAUTILUS events 
(with hourly $\overline{T_{eff}}\leq 10~mK$). We accept only the events
with the ten minute average $T_{eff}~<~7~mK$. Note that the EXPLORER events
are more numerous than the NAUTILUS events because of the different
bandwidth.
        \label{fig3} }
\end{figure}

To this end we consider the quantity $T_{eff}$,
which we used in two ways. The first was to compute 
$T_{eff}$ by averaging $x^2$ over one hour of continuous measurements
($\overline{T_{eff}}$),
the second to consider the $T_{eff}$ averaged during the
ten minutes preceeding each event. For the hourly averages we
show the distribution in fig.\ref{fig2}.
On observing this figure we decided to consider for the search
for coincidences only the time periods with hourly averages
smaller than 10 mK.
The distributions for the $T_{eff}$ averaged over the ten minutes
preceeding each event are shown in fig.\ref{fig3}
for EXPLORER and NAUTILUS.
It will be noticed that the number of events is larger for EXPLORER than
for NAUTILUS. This depends on the bandwidth $\Delta f$ which is larger
for EXPLORER(see Table \ref{dire}).
 
On observing these distributions 
we decided, as conservative $a~priori$ data selection,
to make a cut and accept only the events for which the corresponding
$T_{eff}$ was below $7~mK$ (and there was no seismometer veto).
This meant regarding the bump at 10 mK, for NAUTILUS, as due to 
extra noise. 
We recall that in the previous search \cite{astone2001} with the noisier
1998 data the cuts on $T_{eff}$ were made at between 25 and 100 mK.

From our previous experience we had learned that the detectors
operate in a more stationary way when the noise temperature
remains low for longer periods of time, because this indicates
a smaller contribution of extra noise. To make a quantitative check
on this we classified the data stretches in various categories,
according to the length of the continuous periods having hourly averages
$\overline{T_{eff}}~\leq~10~mK$,
 obtaining the figures shown in Table \ref{cate}.
\begin{table}
\centering
\caption{
In the first column we indicate the minimum time length of continuous
operation, in the second column the total common time of
measurement. The last columns indicate the number of events, only
those during
the common time of operation, with hourly $\overline{T_{eff}}\leq 10~mK$
and the percentage of the
events having the ten minute average $T_{eff}~>~7~mK$.
}
\vskip 0.1 in
\begin{tabular}{|c|c|cc|cc|}
\hline
time length&hours&EXPLORER &&NAUTILUS&\\
&&events&\%&events&\%\\
\hline
$\ge$1 hour&2156&54762&5.9&11252&37\\
$\ge$3 hour&2082&52683&5.0&10887&34\\
$\ge$6 hour&1927&50344&4.1&9939&31\\
$\ge$12 hour&1490&40105&3.2&7268&27\\
\hline
\end{tabular}
\label{cate}
\end{table}
From this table we clearly see that the longer is the time period
of continuous operation with low noise the smaller is the number
of events associated with a noise $T_{eff} \ge7~mK$.

\begin{figure}
 \vspace{9.0cm}
\includegraphics{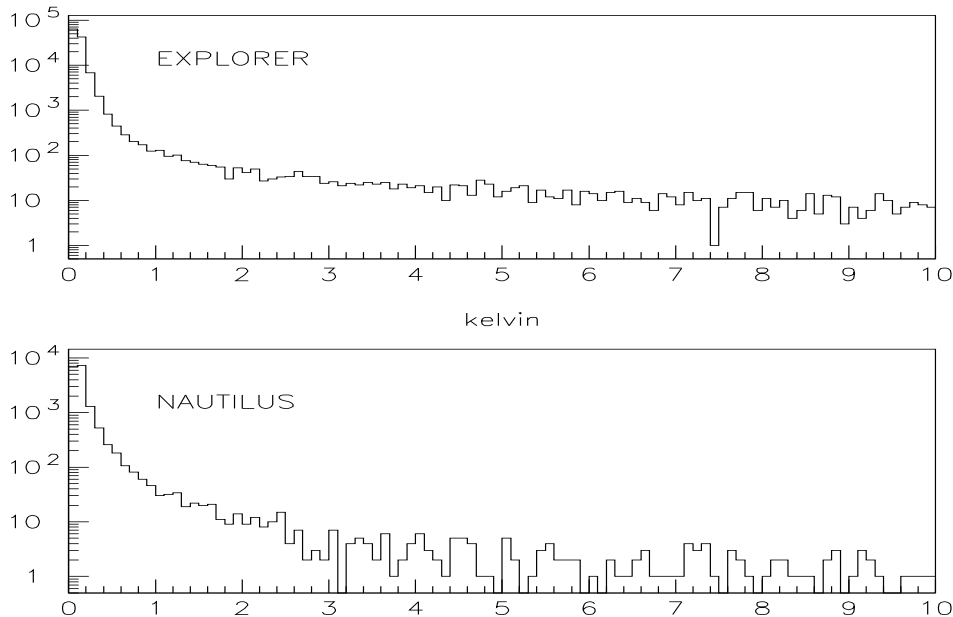}
 \caption{
The distributions of the EXPLORER and NAUTILUS event energies
in kelvin units.
        \label{fig4} }
\end{figure}

Finally, in fig.\ref{fig4} we show the distribution of the event energies
selected according to $\overline{T_{eff}}~\leq~10~mK$ and 
 to $T_{eff} \leq 7~mK$, for each event, belonging to periods with
duration $\geq1~hour$. We notice that, in spite of our selection
criteria, we still have several events with large energy, which indicates
the presence of extra noise, in addition to the thermal and electronic ones.
The only way to eliminate this noise is by means of the coincidence technique.

\section{Searching for coincidences}

For the search for coincidences it is important to establish
the time window.
Using simulated signals and real noise, we characterized \cite{sabrina}
the dispersion of the time of the event around the time
when the signal is applied.

The standard deviation of the time dispersion for a given detector is
\be
\sigma_d=const \frac{1}{\Delta f} \frac{1}{\sqrt{R_e}}
\label{formula}
\ee
for delta signals, 
where $\Delta f$ is the detector bandwidth and the $const=0.28$
is determined, for the EXPLORER and NAUTILUS detectors,
by means of simulation \cite{pianuovo}. For a coincidence analysis
with two detectors we have
\be
\sigma_w=\sqrt{\sigma_{expl}^2+\sigma_{naut}^2}
\label{finestra}
\ee
We decided to take as coincidence window $w=\pm 3 \sigma_w$,
which is the most recent choice of the IGEC collaboration.
Since each event has its own $\sigma_w$ the value of $w$ will be different
for each coincidence. Note that the value of $w$ is almost entirely
due to the
NAUTILUS detector, since EXPLORER has a much larger bandwidth;
it turns out to be of the order of $|w|\sim 0.5~s$, about one half
of that used in previous searches for coincidences. With the use of
$3\sigma_w$ we also take into account the uncertainty of the
detector bandwidth and of the simulation procedure.

Analysis in a coincidence search consists of
comparing the detected number of coincidences at zero time delay ($\pm w$)
with the background, that is with coincidences occurring by chance.
In order to measure the background due to the accidental
coincidences, using a procedure adopted since the beginning of
the gravitational wave experiments \cite{weber},
one shift the time of occurrence of the events of one of
the two detectors a number of times. We shifted 
100 times in steps of $\Delta t=2~s$ (uncorrelated data),
from -100 s to +100 s. For each time shift we get a number of
(shifted) coincidences.
If the time shift is zero we get the number $n_c$ of $observed$ coincidences.
The accidental background is calculated from the average number
of the $n_{shift}$
shifted coincidences obtained from the one hundred time shifts
\be
\bar{n}=\frac{\sum_{j=1}^{100}~n_{shift}(j)}{100}
\label{average}
\ee
This experimental procedure for evaluation of the background
has the benefit of handling 
the problems arising when the distribution of the events
is not stationary (see reference \cite{afp}), although this is not
the case with the present 2001 data. 

\section{Energy filter}

It is clear that if a coincidence between the two detectors is due
to the arrival of a GW burst we expect the energies of the two
coincident events to be correlated, and we can disregard all coincidences
whose corresponding event energies are very different, according to the
considerations illustrated in fig.\ref{fig1}. Thus we can
apply an energy filter with the aim of reducing the background.

The procedure for application of such an energy filter was set 
in our previous search for coincidences \cite{astone2001}.
We considered signals of various energies $E_s$.
For each coincidence found we calculated the
$R_s$ for each of the above signal energy $E_s$ using the known values of the
(local ) $T_{eff}$ of the two events. We then verified whether the
two $R_e$, for the two events of
that coincidence, fell within the interval $R_s \pm \Delta R_s$,
such that the two limits $R_s-\Delta R_s$ and $R_s+\Delta R_s$
delimitate (see fig.\ref{fig1}) an area of 68\%
 (about one standard deviation for well behaved noise) for
a given value of $E_s$; that is, we verified the compatibility
of the two events.
We followed the same procedure for the shifted
coincidences in order to estimate the background after application
of the energy filter.
In this way we reduced, for the 1998 data,
 the average number of accidental coincidences from
$\bar{n}=223$ to $\bar{n}=51$.

This procedure is useful, in particular, if the two detectors have
different sensitivity, as in the case of the 1998 data, and, consequently,
the event thresholds and the event energies
are also different. In the case of two detectors with comparable
sensitivity, however, one could also consider to compare directly the energies
of the coincident events. 
For the 2001 data, although in this year the two detectors had
comparable sensitivities
(see fig.\ref{fig2}), we decided not to change the procedure
used for the 1998 data. We considered GW signals
of energy $E_s$,in a range covering the energies of our events,
i.e. $E_s$ from 5 mK to 1 K in steps of 5 mK, and accepted the coincidence
(at zero delay or at a shifted time) if the two events
fell within the above interval $R_s \pm \Delta R_s$.

\section{Sidereal time distribution}

In our previous search for coincidences \cite{astone2001} we 
took into consideration the non-isotropic
response of the detector to a GW burst. We had reasoned that,
since extragalactic GW signals should not be detected with the present
detectors, possible sources should be located in our Galaxy,
 or in the Local Group. If any of these sources
exist we should expect a more favorable condition of detection
when the detectors are oriented with their axes perpendicular
to the direction of the potential source, since
the bar cross-section is proportional to $sin^4(\theta)$,
where $\theta$ is the angle between
the detector axis and the direction of the line joining it with
 the source. We did find
a small coincidence excess when angle $\theta$ with respect to
the Galactic Centre was larger than
a certain value (see fig.3 of Ref.\cite{astone2001}).
The inconvenience of this method is that, due to the poor statistics,
the result has to be presented in an integral type graph, which makes it
difficult to appreciate the real statistical significance of the data.
Furthermore, hypotheses must be made on the location of the GW source.

In the present search for coincidences we extend the previous analysis
as follows.
We still make use of the directional property of the antenna
cross-section. As the Earth rotates around its axis, during the day
the detector happens to be variably oriented with respect to a given source
at an unknown location. Thus we expect the signal to be modulated
during the day; more precisely the modulation is expected to have
a period of one sidereal day (with one or more maxima)
(see references \cite{webersid,bary}), since the GW sources,
if any, are  certainly located far outside our Solar system.

The principal, key analysis is carried out with the events in the time periods
of at least twelve hours of continuous data taking 
(see Table \ref{cate}) to which the energy filter is applied.

\begin{figure}
 \vspace{9.0cm}
\includegraphics{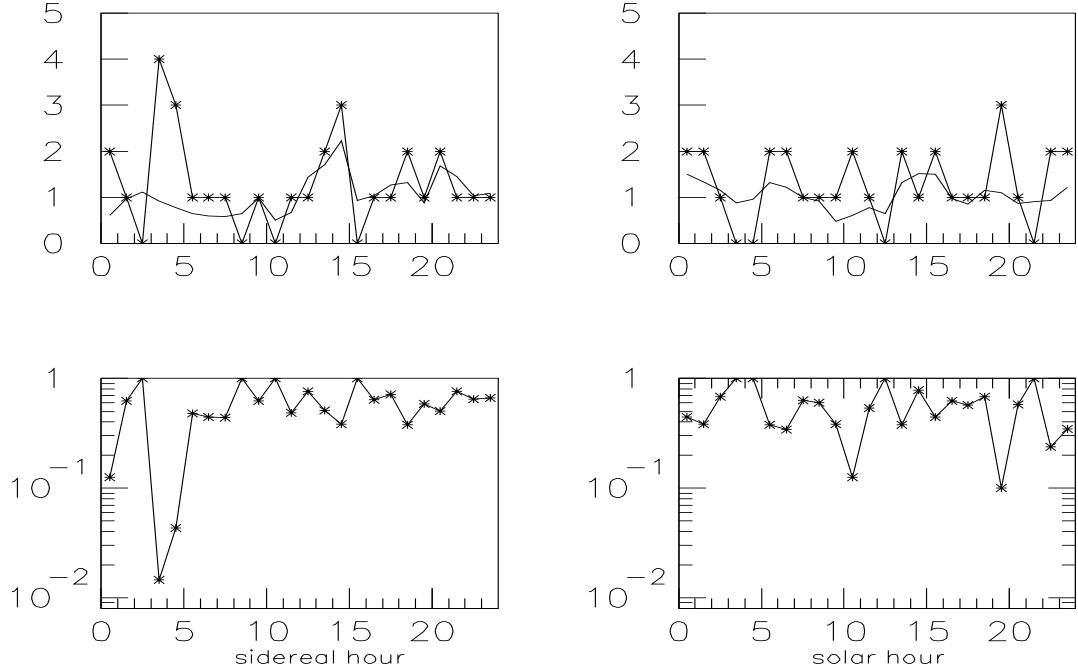}
 \caption{
Result with events in the long time periods ($\ge~12~hour$) of
continuous operation.
The upper graph on the left shows the number of coincidences $n_c$ indicated
with the * and the
average number $\bar{n}$ of accidentals versus the sidereal hour.
The lower graph on the left shows the Poisson
probability to obtain a number of coincidences greater than or equal
to $n_c$.
The two graphs on the right show the result using the solar time in hours.
We remark that the data points refer to independent sets of events.
        \label{fig5} }
\end{figure}

Twenty-four categories of events are considered, one per each sidereal hour,
 the sidereal time referred to
a position and orientation halfway between EXPLORER and
NAUTILUS (this determines the zero local sidereal time which is not
essential for the following considerations).
 Each category includes coincidences totally
independent from those in the other categories.
For each category in fig. \ref{fig5}
we report the number $n_c$ of observed coincidences,
the average number $\bar{n}$ of accidental coincidences
obtained by  using the time shifting procedure
and, given $\bar{n}$, the probability $p$ that a number $\ge n_c$ of
coincidences could have occurred by chance.
For comparison we also show a histogram produced with the same
procedure using solar hours.
 
One notice a coincidence excess from sidereal hour 3 to 
sidereal hour 5, which
appears to have some statistical significance, as the two largest
excesses occur in two neighboring hours (the events in each hour are
totally independent from those in a different hour). We have $n_c=7$
coincidences in this two-hour interval and $\bar{n}=1.7$.
On the contrary, no significant coincidence excess appears at any solar hour.

\begin{figure}
 \vspace{9.0cm}
\includegraphics{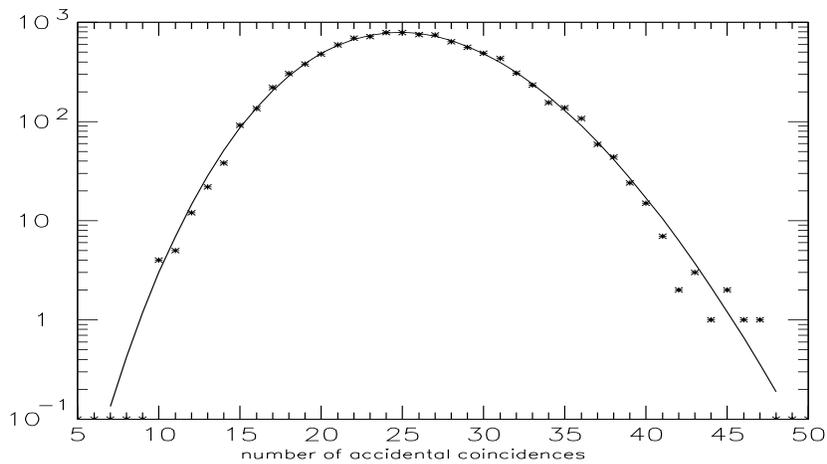}
 \caption{
The distribution of the number of accidental coincidences 
obtained with ten thousand trials, with average
number $\bar{n}=25.32$. The stars indicate the experimental
distribution, the continuous line the expected Poissonian distribution.
        \label{fig6} }
\end{figure}

The accidental coincidences always have a Poissonian
distribution. To check this, we have considered for all the above events
the accidental coincidences obtained with ten thousand trials, 
by time shifting from -10000 s to 
+10000 s in steps of two seconds. The distribution of the number of accidental
coincidences is shown in fig. \ref{fig6}. The agreement between
experimental and expected distributions is excellent.

\begin{figure}
 \vspace{9.0cm}
\includegraphics{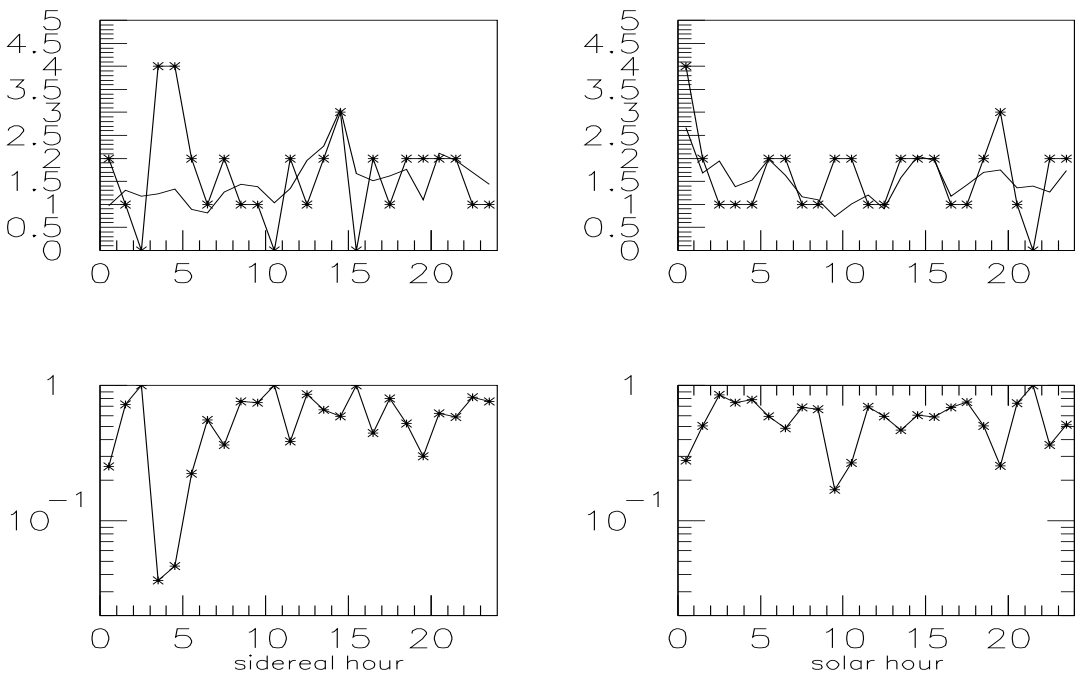}
 \caption{
Result with events in time periods $\ge~1~hour$ of continuous
operation. As in fig.\ref{fig5}. 
        \label{fig7} }
\end{figure}

We repeat the analysis for the events belonging to the
larger set of continuous
data taking lasting one hour or more (first line of Table\ref{cate}).
We obtain the result shown in fig.\ref{fig7}.
We notice that in the sidereal two-hour interval the number of coincident events
has increased from seven to eight and the background has become
$\bar{n}=2.6$.

\begin{figure}
 \vspace{9.0cm}
\includegraphics{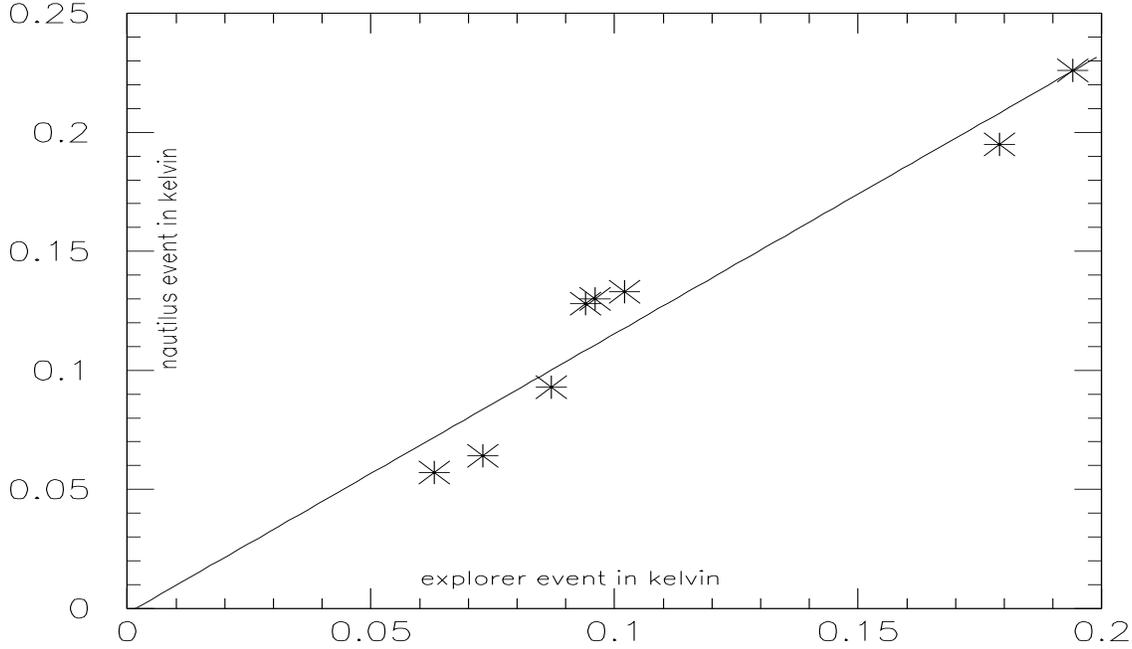}
 \caption{
Correlation between the event energies of NAUTILUS with those of
EXPLORER for the eight coincidences occurred in the sidereal hour
interval 3 to 5, in time periods $\ge~1~hour$.
The correlation coefficient is 0.96. No energy filter was applied.
        \label{fig8} }
\end{figure}

We proceed now in performing a test, to check if this result
is compatible with simultaneous physical excitation of the two
detectors of possible non-terrestrial origin.
We compare the energies of the coincident events: if the events
in EXPLORER and NAUTILUS are due to the same cause we expect
their energies to be correlated. 
This test must be done without applying the energy filter.
Using the events in the time periods with duration $\geq 12~hours$
and without applying the energy filter we still get seven coincidences.
If we consider the time periods with duration $\geq 1~hour$
we get eight coincidences.
The event energies are very strongly correlated, as shown in
fig.\ref{fig8}.
We also studied the
energy correlation of the events of the accidental coincidences and
found no correlation.

\begin{figure}
\vspace{9.0cm}
\includegraphics{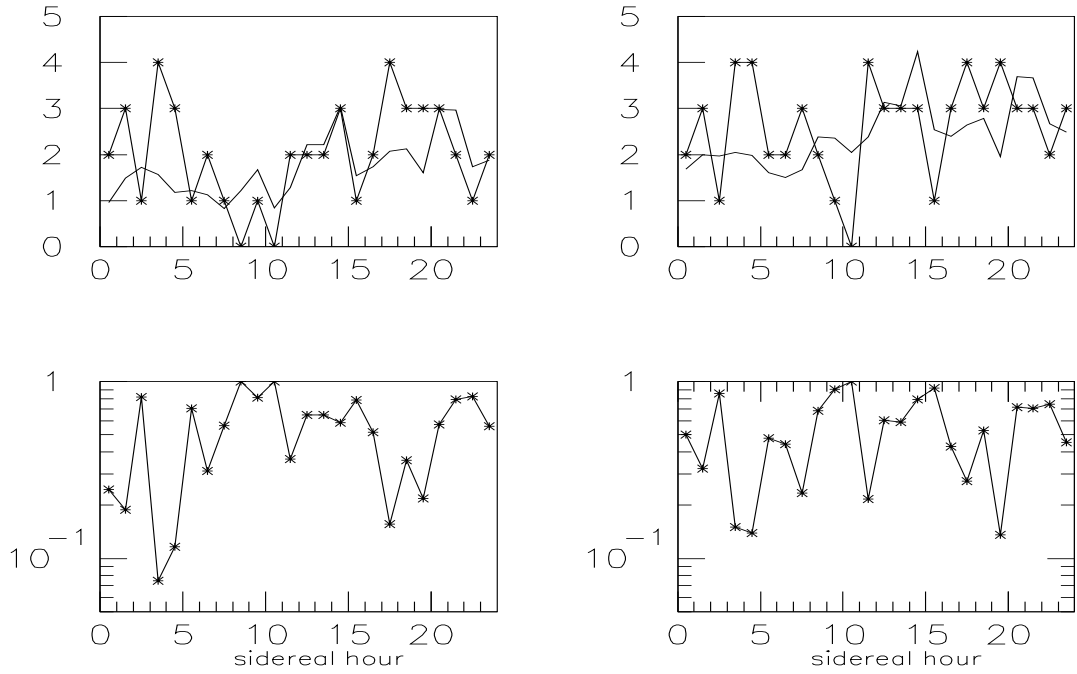}
 \caption{
As in fig.\ref{fig5}, with no application of the energy filter.
The two graphs on the left refer to events in time periods with duration 
$\ge~12~hours$. The two graphs on the right refer to events in time
periods with duration $\ge~1~hour$.
        \label{fig9} }
\end{figure}

We also performed a coincidence data analysis when no energy filter
at all was applied, obviously expecting a larger accidental
background. The result is shown in fig.\ref{fig9}.
We find that the coincidence excess in the time interval 3 to 5
sidereal hours still shows up, although less clearly, as expected.

\begin{figure}
\vspace{9.0cm}
\includegraphics{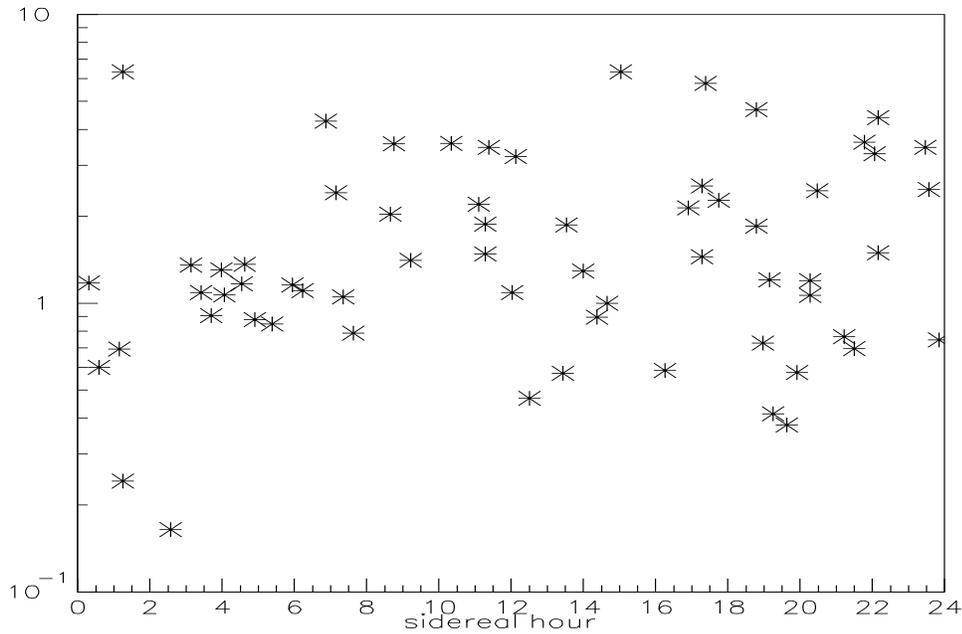}
 \caption{
Ratios of the energies (nautilus/explorer) for the events 
in coincidence belonging to periods with duration $\geq 1~hour$
(no application of the energy filter) versus the sidereal hour.
        \label{fig10} }
\end{figure}

A different way to present these data is shown in fig.\ref{fig10}.
This figure shows that at sidereal hours outside the interval 3 to 5 hours
the event energies are not correlated. Also it shows the particular behaviour
in the 3 to 5 hour interval, when the energies of all coincident events
are correlated.

The eight events in the 3 to 5 hour period are listed in the Table \ref{coinci}.
\begin{table}
\centering
\caption{
List of the coincident events at sidereal hours between 3 and 5.
No energy filter has been applied. $\delta t$ is the time difference
between the two coincident events and $E$ is the event energy.
}
\vskip 0.1 in
\begin{tabular}{|c|c|c|c|c|cc|cc|c|}
\hline
day&hour&min&sec&$\delta t$&EXPLORER&$[mK]$&NAUTILUS&$[mK]$&sidereal\\
&&&&$[s]$&$E$&$T_{eff}$&$E$&$T_{eff}$&hour\\
\hline
112& 13& 59& 33.26&  0.01& 94&3.9& 128&2.9&  4.6\\
130& 11& 39& 28.07& -0.08& 179&3.6& 195&5.9&  3.4\\
133& 12& 35& 45.40& -0.12& 194&7.0& 226&5.7&  4.5\\
166& 10& 48&  6.04&  0.39& 73&3.2& 64&3.0&  4.9\\
198&  7& 46& 40.32&  0.43& 102&2.9& 133&3.6&  4.0\\
278&  2& 12& 29.65&  0.37& 63&2.9& 57&2.6&  3.7\\
296&  0& 29& 40.59& -0.12& 96&2.6& 130&5.6&  3.1\\
296&  1& 24& 10.46&  0.00& 87&2.8& 93&4.1&  4.1\\                                
\hline
\end{tabular}
\label{coinci}
\end{table}
We have verified that, using the cosmic ray detector of NAUTILUS,
these events are not due to cosmic ray showers.

\section{Comparison with the 1998 data}

The analysis presented here differs slightly (eg. for the use of the
sidereal time) from that applied previously \cite{astone2001}.
We therefore present the 1998 data also in terms of the sidereal
time.
We must consider that the 1998 data are noisier than the 2001 data.
In particular the EXPLORER data have a noise $T_{eff}$ ten times
larger than that of the 2001 data, whilst the NAUTILUS noise was
of the same order.

\begin{figure}
 \vspace{9.0cm}
\includegraphics{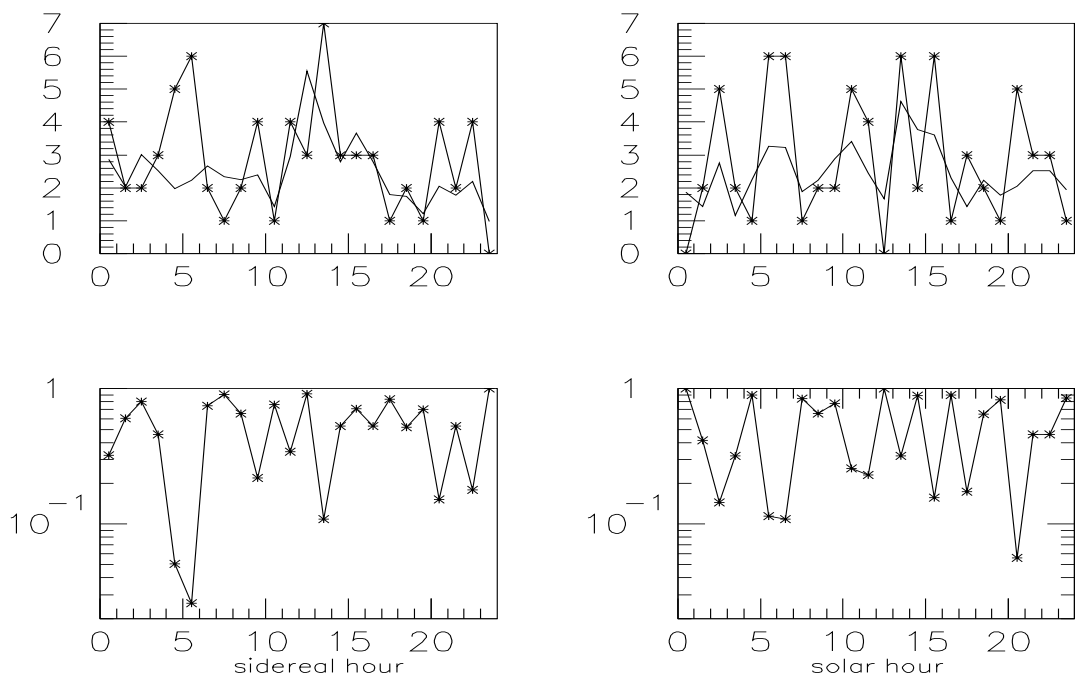}
 \caption{
1998 data with energy filter.
The upper graph on the left shows the number of coincidences $n_c$ indicated
with the * and the
average number $\bar{n}$ of accidentals versus the sidereal hour.
The lower graph on the left shows the Poisson
probability of obtaining a number of coincidences greater or equal
to $n_c$.
The two graphs on the right show the result using the solar time in hours.
        \label{fig11} }
\end{figure}

For the coincidence search we change the window from $w=\pm1~s$
(the IGEC choice at that time) used in the paper \cite{astone2001}
to the present $w=\pm3 \sigma_w$ used here.
The result is given in fig.\ref{fig11}.
An effect similar to the one found for the 2001 data is noticed,
although weaker than that obtained with the less noisy 2001 data.

\section{Robustness of the statistical analysis}
In any data analysis care must be taken to avoid choosing
procedures which favour particular results. Thus we considered
the possibility that our result be biased, although involuntarily,
by any such choices. In the present analysis all choices were
made "a priori" and already published in the 
scientific literature; in particular the IGEC choice for the
coincidence window of $\pm3~\sigma_w$, the energy filter
(see ref. \cite{long} and \cite{astone2001})
and the threshold for definition of an event (see ref.\cite{5barre}).

Nevertheless, we tested whether our present choices were indeed, to accident,
most apt to produce the coincidence excess. 
This proved not to be the case. 

To find the $most~favorable$
parameters, that is the threshold and the energy filter parameter,
we considered only the
coincidences in the sidereal hour interval 3 to 5 and minimized the
probability of a coincidence excess by chance. We found that the  
most favorable threshold for the definition
of event is at $R_t=20.5$, instead of $19.5$.
 The most favorable parameter for the
energy filter is 50\%, instead of 68\%.

\begin{figure}
 \vspace{9.0cm}
\includegraphics{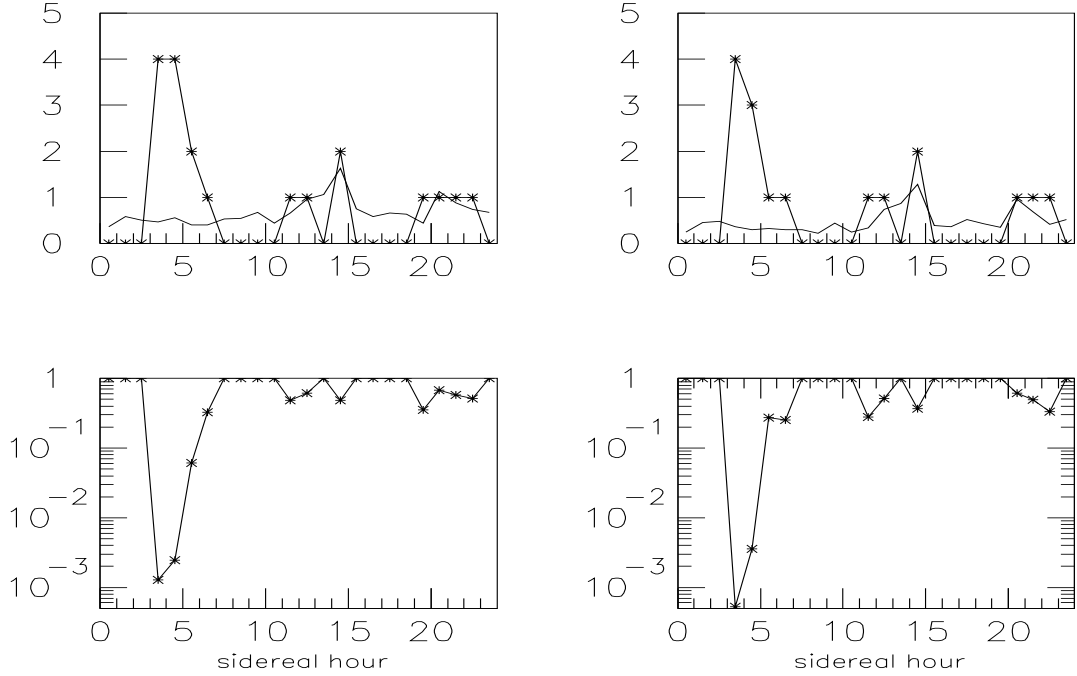}
 \caption{
2001 data with "a posteriori" choices (see text).
 As in fig.\ref{fig7} for the two graphs
on the left (periods $\ge 1~hour$) and as in fig.\ref{fig5}
for the two graphs on the right (periods $\ge 12~hour$).
 Again we remark that the data points in the figure at
different sidereal hours are independent one from the other.
        \label{fig12} }
\end{figure}

We thought it interesting to report the result for these most favorable choices,
$R_t=20.5$ for the threshold and 50\% for the energy filter.
The result is shown in fig.\ref{fig12}.
We notice an indication that the coincidence excess might extend to
the sidereal hour interval 3 to 6, including two more coincidences
in the period 5 to 6 sidereal hours (see also fig.\ref{fig10}).
We want to remark that this
hour was not included in the optimization process.

As far as the coincidence window is concerned, we found ($a~posteriori$)
that the best choice for having a coincidence excess in the
3 to 5 sidereal hour interval is $\pm3.5\sigma_w$, and any coincidence
window from $\pm 2.5\sigma_w$ to $\pm 4\sigma_w$ gives comparable results.

\section{Discussion and conclusions}

The IGEC search for coincidences \cite{5barre} was performed without applying 
the event energy algorithms based on the event amplitude and on the directional
properties of the detectors; it gave a no coincidence excess.
With new data taken in the year 2001 and with improved
sensitivity we repeated the coincidence search with the detectors
EXPLORER and NAUTILUS (no other detector was in operation during the
year 2001), applying data analysis algorithms
based on known physical characteristics of the detectors, namely energy
of the events and directionality of the detectors. 
We obtained a coincidence excess at sidereal hours between 3 and 5.

\begin{figure}
 \vspace{9.0cm}
\includegraphics{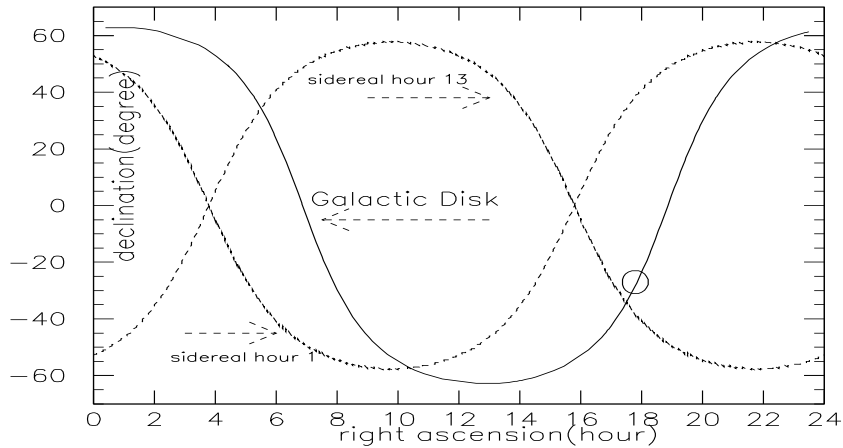}
 \caption{
The position of the Galactic Disk and
the loci of point sources perpendicular to the detector axis at 1
and 13 sidereal hour in the right ascension-declination plane.
At sidereal hour 4.3 the locus tends to coincide with
the line of the Galactic Disk. The large circle indicates the location
of the Galactic Centre.
        \label{striscia-ora} }
\end{figure}

At a given sidereal time the intersection of the celestial sphere
with the plane perpendicular to the detector axis is a circle.
We show in fig.\ref{striscia-ora} two of these circles (i.e., at
1 and at 13 sidereal hours) in the right ascension-declination plane.
In the representation of fig.\ref{striscia-ora}
the line which indicates the point sources perpendicular
to the detector axis (we call this $~the~line~of~maximum~sensitivity$)
moves to the right with sidereal time.
This line intersects the location of the Galactic Centre twice a day
(at 4.3 and at 13.6 sidereal hours). Only once per day (at 4.3 sidereal hour)
this line overlaps with the entire Galactic Disk.
The overlapping takes place because of the particular orientation of
the detectors on the Earth' surface (see Table \ref{dire}), and would
not occur for a different azimuth angle of detector orientation.

If all GW sources were concentrated in the Galactic Centre we would have
found a coincidence excess twice a day.
The coincidence excess occurs only once per day, just when the line of maximum
sensitivity of the detectors overlaps with the Galactic Disk, as if GW sources
were distributed in the Galactic Disk and not just located in its Center.

As for the energy balance, in the year 2001 we find in the
interval from 3 to 5 sidereal hour
a coincidence excess of, very roughly, $n_c-\bar{n}\sim 6$ 
coincidences occurring in five days (two sidereal hours out of twentyfour, in a
total  time period of 1490 hours $\sim$60 days).
In terms of energy conversion
into GW we have, very roughly, about one coincidence
per day with a signal energy of about 100 mK. This corresponds,
using the classical cross-section, to a conventional burst with amplitude
$h\sim 2~10^{-18}$ and to the isotropic conversion into GW energy of 0.004
solar masses, with sources located at distance of 8 kpc.
The observed rate is much larger than the models today available predict,
for galactic sources. We note, however, that our rate of events is within
the upper limit determined by IGEC \cite{5barre} for short GW bursts
and by the 40m-LIGO prototype interferometer
\cite{ligo} for coalescing binary sources in the Galaxy.

We think it is unlikely that the observed coincidence
excess be due to noise fluctuations, but we prefer to take a conservative
position and wait for a stronger confirmation of our result, before reaching
any definite conclusion and claim
that gravitational waves have been observed. Furthermore, although we have 
excluded that the events are due to cosmic ray showers 
(see section 6), we cannot
completely rule out that they be due to some other exotic, still unknown,
phenomenon. A possible way to distinguish GW from other causes is to
measure other vibrational modes of the detectors and verify that,
as predicted by General Relativity, only quadrupole modes are excited.
 This requires multimode detection (with bars or spheres) and to improve
 the signal-to-noise ratio of the apparatuses.

We expect to collect new data with EXPLORER and NAUTILUS
with improved sensitivity. We plan to repeat the
same analysis with these new data and also with any other new data
provided by other GW groups, those which operate the resonant detectors and 
those which operate or are beginning to operate the
interferometric detectors GEO, LIGO, TAMA and VIRGO.

\section{Acknowledgements}

We are indebted to Professor Ugo Amaldi for very useful
discussions and suggestions.
We thank the CERN cryogenic facility and
F. Campolungo, R. Lenci, G. Martinelli, A.Martini, E. Serrani, R. Simonetti 
and F. Tabacchioni for precious technical assistance.

%
%

%
\end{document}